\documentclass[english,floatfix,twocolumn,superscriptaddress,nofootinbib,aps,prd]{revtex4}
\usepackage[utf8]{inputenc}
\usepackage[T1]{fontenc}
\usepackage{lmodern}
\usepackage{amsmath}
\usepackage{amssymb}
\usepackage{graphicx}
\usepackage{esint}
\usepackage{longtable}
\usepackage{dcolumn}
\usepackage{babel} 
\usepackage{csquotes} 
\DeclareMathOperator{\Ima}{Im}

\begin{document}

\title{Bardeen regular black hole as a quantum-corrected Schwarzschild black hole}
\author{R. V. Maluf}
\email{r.v.maluf@fisica.ufc.br}
\affiliation{Universidade Federal do Ceará (UFC), Departamento de Física,\\ Campus do Pici, Fortaleza,  CE, C.P. 6030, 60455-760 - Brazil.}

%%%%%%%%%%%%%%%%%%%%%%%%%%%%%%%%%%%%%%%%%%%%%%%%%%%%%%%%%%%%%%%%%%%%%%

\author{Juliano C. S. Neves}
\email{nevesjcs@if.usp.br}
\affiliation{Centro de Ciências Naturais e Humanas, Universidade Federal do ABC,\\ Avenida dos Estados 5001, Santo André, 09210-580 São Paulo, Brazil}

\begin{abstract}
Bardeen regular black hole is commonly considered as a solution of general relativity coupled to a nonlinear electrodynamics. In this paper, it is shown that the Bardeen solution may be interpreted as a quantum-corrected Schwarzschild black hole. This new interpretation is obtained by means of a generalized uncertainty principle applied to the Hawking temperature. Moreover, using the regular black hole of Bardeen, it is possible to evaluate the quantum gravity parameter of  the generalized uncertainty principle or, assuming the recent upper bounds for such a parameter, to verify an enormous discrepancy between a cosmological constant and that measured by recent cosmological observations $(\sim 10^{120})$.
\end{abstract}

\maketitle

\section{Introduction}

Bardeen's regular metric \cite{Bardeen2} is considered the first regular black hole (RBH) ever constructed  in general relativity (GR). It is defined as a spherical object of variable mass, $m(r)$, with horizon(s) (at most two: the inner $r_-$ and the outer $r_+$ horizons) and without the physical singularity. Such a regular geometry is possible because, according to ideas of Sakharov and others \cite{Sakharov,Gliner}, spacetime is de Sitter-like in regions of matter with high density. For RBHs, there exists a de Sitter core for small values of the radial coordinate, i.e., inside the event horizon. With a de Sitter core, Bardeen RBH violates the strong energy condition. The violation of energy conditions is the origin for the regularity of RBHs \cite{Neves_Saa}, thereby RBHs do not obey at least one condition of Hawking-Penrose's theorems of singularity \cite{Wald_book}. Then, for RBHs, the existence of a singular point (or a singular ring for axisymmetric spacetimes) is not a necessary consequence from the theorems of singularity. 

In this paper, we will interpret the Bardeen RBH from another point of view. Considering the general metric with spherical symmetry\footnote{We adopt Planck units, where the speed of light in vacuum $c$, the gravitational constant $G$, Boltzmann constant $k_B$, and the reduced Planck constant $\hslash$ are set equal to 1. In some calculations, $c$, $G$, and $\hslash$ are restored.}
\begin{equation}
ds^2=-f(r)dt^2+\frac{dr^2}{f(r)}+r^2 \left( d\theta^2+\sin^2\theta d\phi^2 \right),
\label{Metric}
\end{equation}
where $f(r)=1-2m(r)/r$, the Bardeen solution appears due to a suitable choice for $m(r)$. Following the notation of Neves and Saa \cite{Neves_Saa}, the mass function for the Bardeen RBH is written as
\begin{equation}
m(r)= M \left[1+\left(\frac{r_0}{r}\right)^{2} \right]^{-\frac{3}{2}},
\label{Mass}
\end{equation}
where $M$ and $r_{0}$ may be interpreted as mass and length parameters, respectively. The parameter $M$ stands for the ADM mass of Schwarzschild black hole. The limit of $m(r)$ confirms such a interpretation, i.e., $\lim_{r\rightarrow\infty}m(r)=M$. Indeed, the metric (\ref{Metric}) with the mass function $m(r)$ is approximately the Schwarzschild metric for large values of the radial coordinate ($r \gtrsim r_+$). On the other hand, contrary to Ayon-Beato and Garcia's interpretation \cite{Beato}, according to which $r_0=e$ is a magnetic monopole, and the Bardeen solution comes from GR coupled to a nonlinear electrodynamics, we consider $r_0$ as a microscopical length, and, as we will see, such a length parameter is proportional to the Planck length. Without such a microscopical length ($r_0=0$), the metric (\ref{Metric}) with the mass function (\ref{Mass}) is the Schwarzschild metric. The most interesting feature provided by the mass function given by Eq. (\ref{Mass}) is a de Sitter core for small values of $r$.  For these scales, $m(r) \sim r^3$, and the metric term is $f(r)\sim 1-Cr^2$, with $C$ playing the role of a positive cosmological constant.

Thermodynamics of black holes is one of the most interesting chapters in black hole physics. The concepts of temperature, entropy and heat capacity, for example, are associated to black holes as well. Classically, black holes do not emit radiation. However, from a semiclassical point of view, Hawking \cite{Hawking} showed that black holes emit radiation. Today it is called Hawking radiation and obtained from several ways. Besides the original Hawking's approach, the tunneling method provides a black hole radiation. In such an approach, a particle may cross the event horizon by quantum tunneling. It is possible to indicate two ways to obtain the tunneling result in the literature: the first one is the null-geodesic method developed by Parikh and Wilczek \cite{Parikh}; the second one---constructed by Agheben \textit{et al.} \cite{Agheben}---uses the Hamilton-Jacobi ansatz. The latter will be used in this paper to evaluate both the Hawking temperature and the quantum-corrected Hawking temperature. 

The tunneling method will be useful to calculate the quantum-corrected temperature using a generalized uncertainty principle (GUP).\footnote{See Tawfik and Diab \cite{Tawfik} for a review on GUP and its applications.} A GUP provides high energy corrections to black holes thermodynamics, which come from both a quantum theory of gravity and the idea of a minimum length. With a specific GUP applied to the quantum tunneling, we will interpret the Bardeen RBH as a quantum-corrected Schwarzschild black hole at second order of approximation. 

The structure of this paper is as follows. Sec. 2 presents an application of the tunneling approach to the Bardeen RBH temperature. Sec. 3 shows a GUP used for the Bardeen RBH, and Sec. 4, with the aid of the results of the previous section, indicates that such a regular solution may be thought of as a quantum-corrected Schwarzschild black hole. Final remarks are presented in Sec. 5.

\section{Tunneling approach}

As we said, the quantum tunneling effect allows that particles inside the black hole cross the event horizon. It is possible to calculate the tunneling probability of this process. In such a method, we are interested in radial trajectories. Therefore, the metric (\ref{Metric}) may be considered 2-dimensional near the horizon:
\begin{equation}
ds^2=-f(r)dt^2+\frac{dr^2}{f(r)}.
\label{metric-2d}
\end{equation}
Thus, the problem is entirely solved in the $t-r$ plane. The Klein-Gordon equation for a scalar field $\phi$ with mass $m_{\phi}$, with the aid of Eq. (\ref{metric-2d}), is written as
\begin{equation}
-\partial_{t}^{2}\phi +f(r)^2 \partial_{r}^{2}\phi+\frac{1}{2}\partial_{r}f(r)^2\partial_{r}\phi-\frac{m_{\phi}^2}{\hslash}f(r)\phi=0.
\label{equation}
\end{equation}
Now, using the WKB method, one has the following solution for Eq. (\ref{equation}):
\begin{equation}
\phi (t,r)= \exp \left[-\frac{i}{\hslash}I(t,r) \right].
\label{phi}
\end{equation}
And for the lowest order in $\hslash$, substituting Eq. (\ref{phi}) into Eq. (\ref{equation}), we have the so-called Hamilton-Jacobi equation
\begin{equation}
\left(\partial_{t}I \right)^2 -f(r)^2 \left(\partial_{r}I \right)^2-m_{\phi}^2f(r)=0,
\label{HJ}
\end{equation}
with the action that generates (\ref{HJ}) given by
\begin{equation}
I(t,r)=-Et+W(r).
\label{Action}
\end{equation}
$E$ is the radiation (or particle) energy. As we will see, $E$ is the energy of the Hawking radiation. The explicit form for $W(r)$, the spatial part of the action, reads 
\begin{equation}
W_{\pm}(r)=\pm \int\frac{dr}{f(r)}\sqrt{E^2-m_{\phi}^2f(r)}.
\label{W}
\end{equation}
The functions $W_{+}(r)$ and $W_{-}(r)$ represent outgoing and ingoing solutions, respectively. It is important to emphasize that classically $W_{+}(r)$ is forbidden because it describes solutions that cross the event horizon, moving away from $r_+$. Then, to obtain the Hawking radiation outside the event horizon we will focus on $W_{+}(r)$. 

With the approximation for the function $f(r)$ near the event horizon, i.e., $f(r)= f(r_+)+f'(r_+)(r-r_+)+...,$ Eq. (\ref{W}) assumes the simple form:
\begin{equation}
W_{+}(r)=\frac{2\pi i E}{f'(r_+)},
\end{equation}
with $'$ denoting derivative with respect to $r$. Thus, the probability of crossing the event horizon by tunneling, or the emission rate, is given by the imaginary part of the action (\ref{Action}):
\begin{equation}
\Gamma \simeq \exp \left[-2 \Ima I \right] = \exp\left[-\frac{4 \pi E}{f'(r_+)} \right].
\label{Gamma}
\end{equation}
Comparing Eq. (\ref{Gamma}) with the Boltzmann factor, namely $e^{-E/T}$, the Hawking temperature derived by the tunneling method is written as
\begin{equation}
T=\frac{E}{2\Ima I}=\frac{f'(r_+)}{4\pi}.
\label{Tt}
\end{equation} 
As we can see, the temperature obtained by tunneling is the same obtained by the surface gravity $\kappa$ used by Hawking. It is rather interesting to obtain Bardeen temperature using both Eq. (\ref{Tt}) and an approximate mass function. For large values of $r$ (specifically for $r\gtrsim r_+$ or, equivalently, $r_0/r \ll 1$), the mass function is $m(r)\approx M \left(1-\frac{3}{2}\left(\frac{r_0}{r} \right)^2 \right)$, and the metric term in the Bardeen RBH reads 
\begin{equation}
f_{B}(r) \approx 1-\frac{2M}{r}\left[ 1- \frac{3}{2} \left(\frac{r_{0}}{r}\right)^2 \right],
\label{Approx}
\end{equation}
up to second order in $r_0/r$. This approximation leads to a simplified result for temperature of the Bardeen RBH and, as we will see, a clear comparison with the quantum-corrected temperature of Schwarzschild black hole. Substituting the metric term (\ref{Approx}) into Eq. (\ref{Tt}), the approximate temperature of Bardeen RBH is given by 
\begin{equation}
 T_{B}  \approx T_{Sch} \left[1- \frac{9}{2} \left(\frac{r_{0}}{r_+}\right)^2 \right] ,
\label{Approximate}
\end{equation}
in the same order of approximation for $r_0/r_+$, with $ T_{Sch}=1/4\pi r_+$  playing the role of the Schwarzschild temperature, and $r_+$ is approximately the Schwarzschild radius. Therefore, the approximate temperature of the Bardeen RBH is the temperature of Schwarzschild black hole minus a positive term from the regular metric. Then, according to Eq. (\ref{Approximate}), \textit{the Bardeen RBH is colder than the Schwarzschild black hole}. This interesting feature may be conceived of as an observational difference between regular and singular black holes.\footnote{Indicated by us \cite{Maluf_Neves} from a general class of RBHs.}

\section{A generalized uncertainty principle}

The GUPs (there are generalized principles \cite{Tawfik}) have been applied in several black holes metrics. Besides the standard black holes \cite{Various_GUP,Various_GUP2,Various_GUP3} and RBHs \cite{Maluf_Neves} in GR, a GUP was applied in black holes within another context \cite{Anacleto}. 

To obtain quantum corrections to temperature, we will consider a quadratic GUP written as
\begin{equation}
\triangle x\triangle p\geq\hslash\left(1+\frac{\lambda^{2}l_{p}^{2}}{\hslash^{2}}\triangle p^{2}\right),
\end{equation}
where $l_{p}=\sqrt{\frac{\hslash G}{c^{3}}}\approx10^{-35}\ \text{m}$ is the Planck length, and $\lambda$ is the so-called dimensionless quantum gravity parameter. For $\lambda\rightarrow 0$ we have the standard uncertainty relation and, as we will see, the results of the semiclassical approach to obtain the Hawking temperature. We can rewrite the GUP as
\begin{equation}
\triangle p\geq\frac{\hslash \triangle x}{2\lambda^{2}l_{p}^{2}}\left(1-\sqrt{1-\frac{4\lambda^{2}l_{p}^{2}}{\triangle x^2}}\right),
\end{equation}
and, considering $l_p/\triangle x \ll 1$, we apply the Taylor expansion such that
\begin{equation}
\triangle p\geq\frac{1}{\triangle x}\left(1+\frac{2\lambda^{2}l_{p}^{2}}{\triangle x^{2}}+\cdots\right),
\end{equation}
where $\hslash$ was set equal to 1, according to Planck units. By using the saturated form of the uncertainty principle, namely $E\triangle x\geq1$, the energy correction reads 
\begin{equation}
E_{GUP}\geq E\left(1+\frac{2\lambda^{2}l_{p}^{2}}{\triangle x^{2}}\right),
\end{equation}
up to second order in $l_{p}$. Therefore, according to Eq. (\ref{Gamma}), for particles with corrected energy $E_{GUP}$, the tunneling probability of crossing the event horizon is
\begin{equation}
\Gamma_{GUP} \simeq \exp \left[-2 \Ima I_{GUP} \right] = \exp \left[-\frac{4\pi E_{GUP}}{f'(r_{+})}\right].
\label{Gamma_GUP}
\end{equation}
Thus, we may compare  Eq. (\ref{Gamma_GUP}) with the Boltzmann distribution $e^{-E/T}$ to find the quantum-corrected Hawking temperature for particles with energy $E$:
\begin{equation}
T_{GUP}=T \left(1+\frac{\lambda^{2}l_{p}^{2}}{2r_{+}^{2}}\right)^{-1},
\label{T_corrected}
\end{equation}
where $T$ is given by Eq. (\ref{Tt}) and, according to Medved and Vagenas \cite{Vagenas}, the uncertainty in $x$ for events near the event horizon is given by $\triangle x\simeq 2r_{+}$. 
 
\section{Bardeen RBH as a quantum-corrected black hole} 

Let us obtain the quantum-corrected temperature of the Schwarzschild black hole and compare with the temperature of the Bardeen RBH. From the GUP, specifically Eq. (\ref{T_corrected}), assuming that $l_p/ r_+ \ll 1$, one has for the Schwarzschild metric
\begin{equation}
T_{GUP (Sch)} \approx T_{Sch} \left[1- \frac{1}{2} \left(\frac{\lambda l_{p}}{r_+}\right)^2 \right], 
\label{TGUP_Sch}
\end{equation}
up to second order of approximation in $l_p/ r_+$. Such a temperature for particles with energy $E$ is also generated by a metric term in the form
\begin{equation}
f_{GUP(Sch)}(r) \approx 1-\frac{2M}{r}  \left[1- \frac{1}{6} \left(\frac{\lambda l_{p}}{r}\right)^2 \right]. 
\label{f_GUP}
\end{equation}
Assuming and imposing that the microscopical parameter $r_0$ of the Bardeen metric is written as
\begin{equation}
r_0=\frac{\lambda l_p}{3},
\label{R0}
\end{equation}
thus, up to second order in $r_0/r$, or $l_p/r_+$, the metric function for the quantum-corrected Schwarzshild black hole (\ref{f_GUP}) is equal to the metric term for the Bardeen RBH, given by Eq. (\ref{Approx}). Geometrically speaking, the behavior of the metric term for large values of $r$ and the asymptotic limit ($\lim_{r\rightarrow \infty} f(r)=1$) are the same for both black holes. Then, from a thermodynamic point of view, in second order of approximation, both black holes are indistinguishable. 

As we saw, Eq. (\ref{Approximate}) will correspond to the quantum-corrected Schwarzschild temperature, given by Eq. (\ref{TGUP_Sch}), if we assume the relation between $r_0$ and Planck length indicated by Eq. (\ref{R0}). According to Das and Vagenas \cite{Das}, where upper bounds for the quantum gravity parameter were investigated, one has $\lambda < 10^{10}$ (or following the authors notation $\lambda^2 \simeq \beta_{0} < 10^{21}$) using non-gravitational alternatives like the Lamb shift, Landau levels and scanning tunneling microscope. On the other hand, from a gravitational system, specifically a modified Schwarzschild black hole, using values of light deflection and perihelion precession, Scardigli and Casadio \cite{Casadio} obtained worse values, for example $\lambda < 10^{34}$. However, assuming a better upper bound ($\lambda < 10^{10}$), our interpretation of $r_0$ as a microscopical length is confirmed 
\begin{equation}
r_0 < 10^{-25}\ \text{m},
\end{equation}
and, from a specific GUP applied to the tunneling method, we suggest another origin for the Bardeen RBH. Instead of a solution of Einstein's field equations coupled to a nonlinear electrodynamics \cite{Beato}, \textit{we propose Bardeen RBHs as a quantum-corrected Schwarzschild black hole at second order of approximation}. And one of the most important features of potential quantum-corrected black holes is the absence of singularities. As is known, the Bardeen RBH possesses such a feature. 

As we said, the ability of the Bardeen RBH of removing the singularity comes from a de Sitter core inside the event horizon. For small values of $r$, the metric term in Bardeen is written as 
\begin{equation}
f_{B}(r) \approx 1-\frac{2GM}{c^2 r_{0}^{3}}r^2,
\label{Small_r}
\end{equation}
where the physical constants were restored. On the other hand, de Sitter spacetime presents
\begin{equation}
f_{dS}(r)=1-\frac{\Lambda}{3}r^2,
\label{de Sitter}
\end{equation} 
with $\Lambda=\alpha \Lambda_{e}$ playing the role of a positive cosmological constant, and the value of $\Lambda_{e}$, the effective cosmological constant, is given by the recent observations and dark energy data \cite{Planck}, namely $\Lambda_{e}=8\pi G\rho_{\Lambda}/c^4$, where $\rho_{\Lambda}^{m} \sim 10^{-48}\ \text{GeV}^4$ is the measured dark energy density. The dimensionless constant $\alpha$ provides a ratio between both a cosmological constant and the measured cosmological constant. As we will see, recent upper bounds on $\lambda$ may indicate $\alpha \neq 1$.

The comparison between Eqs. (\ref{Small_r}) and (\ref{de Sitter}) provides an interesting relation among several physical constants: 
\begin{equation}
\frac{\Lambda}{3} \simeq \frac{2GM}{c^2 r_0^3}.
\end{equation}
Then, the value of the quantum gravity parameter may be obtained from a RBH using the relation between $r_0$ and $l_p$, shown in Eq. (\ref{R0}), and the above equation. That is,
\begin{equation}
\lambda \sim \frac{1}{l_p}\left( \frac{GM}{\Lambda c^2} \right)^{\frac{1}{3}}.
\label{lambda}
\end{equation}
For a Bardeen RBH with solar mass, $M=M_{\odot}$, we have $\lambda \sim 10^{53}$, considering that the cosmological constant is equal to the effective one ($\alpha=1$). As we said, for a deformed Schwarzschild black hole with solar mass, the best upper bound was $\lambda \sim 10^{34}$ in the paper of Scardigli and Casadio \cite{Casadio}. We argue that a possible reason for this enormous difference (between $\lambda \sim 10^{53}$ and $\lambda \sim 10^{34}$) could be the value of the cosmological constant. Assuming that $\alpha \neq 1$, our values of $\lambda$ are improved. For example, for an extreme $\alpha\sim 10^{120}$, we obtain an interesting $\lambda \sim 10^{13}$, which is closer to the non-gravitational upper bounds indicated by Das and Vagenas \cite{Das}. But such a ratio is the reported discrepancy between the cosmological constant, or vacuum energy, predicted by quantum field theory ($\rho_{\Lambda}^{QFT} \sim 10^{72}\ \text{GeV}^4$), and the measured cosmological constant \cite{Carroll}. Thus, using a recent upper bound for the quantum gravity parameter, our $\alpha$ indicates the ratio---or discrepancy---between a cosmological constant and that measured by the most advanced telescopes.  

\section{Final remarks}

Quantum corrections to the Hawking temperature were studied using a GUP. The GUP applied to the tunneling method to obtain quantum-corrected temperature indicates that the Bardeen RBH may be interpreted as a quantum-corrected Schwarzschild black hole at second order of approximation. Moreover, RBHs, or the Bardeen regular metric, may provide a form to estimate the quantum gravity parameter in the GUP or, from a reliable value for such a parameter, may indicate a cosmological constant in black hole physics that is different from the observed one in cosmology.  

\section*{Acknowledgments}
RVM would like to thank Conselho Nacional de Desenvolvimento Científico e Tecnológico (CNPq) for financial support (Grant No. 305678/2015-9). JCSN would like to thank Department of Physics, at the Universidade Federal do Ceará, for the kind hospitality. This study was financed in part by the Coordenação de Aperfeiçoamento de Pessoal de Nível Superior---Brasil (CAPES)---Finance Code 001.

\end{document}